\documentclass{elsart}

\usepackage{amssymb}
\usepackage{graphicx}

\begin{document}

\begin{frontmatter}

\title{A non subjective approach to the GP algorithm 
for analysing  noisy time series}

\author[label1]{K. P. Harikrishnan} 
\ead{$kp_{\_}hk2002@yahoo.co.in$}
\author[label2]{R. Misra\corauthref{cor1}}
\author[label3]{G. Ambika} 
\author[label2]{A. K. Kembhavi}

\corauth[cor1]{Corresponding author: Address: Inter-University Center for Astronomy and Astrophysics, Post Bag 4,
Ganeshkhind, Pune-411007, India; Phone no: 91-20-25604100; Fax no: 91-20-25604699; Email: rmisra@iucaa.ernet.in }

\address[label1]{Dept. of Physics, The Cochin College,  Cochin-682002, India}
\address[label2]{Inter-University Center for Astronomy and Astrophysics, Post Bag 4,
Ganeshkhind, Pune-411007, India}
\address[label3]{Dept. of Physics, Maharajas College,  Cochin-682011, India}

\begin{abstract}
We present an adaptation of the standard Grassberger-Proccacia (GP) algorithm 
for 
estimating the Correlation Dimension of a time series in a non subjective 
manner. The validity and accuracy of this 
approach is tested using different types of time series, such as, those from 
standard chaotic systems, pure white and colored noise and chaotic systems 
added with noise. The effectiveness of the scheme in analysing noisy time series, 
particularly those involving colored noise, is investigated. An interesting 
result we have obtained is that, for the same percentage of noise addition, 
data with colored noise is more distinguishable from the corresponding 
surrogates, than data with white noise.
As examples for real life applications, 
analysis of data from an astrophysical X-ray object and human brain EEG,  
are presented.
\end{abstract}

\end{frontmatter}

\section{ Introduction}

Most of the complex phenomena observed in nature arise from the nonlinear 
nature of the underlying dynamics. Hence techniques from 
nonlinear dynamics and chaos theory are increasingly being applied to 
diverse fields such as biology and economics,  where finite time series
data of one or two variables are used, even when the model system remains 
unknown. 
One of the most widely used methods to quantify the 
nature of such data,  is the calculation of the correlation dimension $D_{2}$ 
by the Grassberger-Proccacia (GP) algorithm [1,2]. 
Here the scalar time series is used to reconstruct the
dynamics in an embedding space of dimension $M$ using delay coordinates
scanned at a prescribed time delay $\tau$. The results of the analysis are
often  useful to indicate whether the nature of complexity in the data
is chaotic, stochastic or a mixture of the two. 

Such studies have been carried out in various disciplines with beneficial 
outcomes in understanding complex systems [3-5]. 
As examples, we mention the calculation of
the correlation dimension from the light curve of variable stars [6],  
quantifying and predicting the changes in the weather over a short 
period [7],  
analysis of market variables to predict the financial market [8],  
estimation of 
the dimension of the galactic structure in the visible universe [9] etc.  
The dimension 
values calculated from EEG data of the human brain has helped to 
analyze 
various states of the brain and its possible pathological changes [10]. 
The chaotic behavior of erythrocytes deformations identified through this
technique in 
healthy people and dislipidemic patients helps in their treatments.

In all such applications,  the analysis is hampered by the finite 
length of the available data [11] and more importantly by the presence
of noise. The former results in ``edge effects'', producing
a downward bias in the correlation exponent estimates [12,13].
The presence of Gaussian white noise tends to fill the 
available phase space and hence increases the computed $D_{2}$ 
value of the system. This effect has been studied by
many authors [14-16] and it has been shown that reasonable estimation
of $D_2 (M)$ is possible only for moderate noise
contamination. However, white noise, which is characterized
by a flat power spectrum,  is a particular case of the various
kinds of noise that can exist in a physical system.  More problematic is the 
presence of colored noise that  is common in a wide variety of physical 
and biological systems, for example, in Brownian motion,  
astrophysical systems, neurons and in solid state devices. This is so because
pure colored noise  also produces a well saturated value of $D_{2}$ [17]. 
However, colored noise is basically a correlated stochastic process, which 
essentially generates a \emph{random fractal curve} rather than a 
\emph{fractal attractor}, with a power law dependence  
$P(f) \propto 1/f^\alpha$. The power spectral indices $\alpha = 1$ and $2$ 
correspond to ``pink'' and ``red'' noise respectively, while 
$\alpha \rightarrow 0$ gives white noise. Its $D_2$ value depends on   
$\alpha$ by the relation,  $D_{2}= 2/(\alpha -1)$ [17]. 
A detailed analytic study of colored noise has been undertaken by
Theiler[18], where the scaling of the correlation integral for
various values of length scales are derived. 
Also, the effect of  colored noise contamination on the computation of the
correlation dimension has been studied recently by Redaelli et.al. [19], who 
show that the increase in the correlation dimension is less for colored
noise as compared to white noise. The scaling region, which is
used to compute the correlation dimension is less affected
by the contamination of colored noise.

An important aspect of these results is that, since colored noise has
a well defined saturated correlation dimension, 
further analysis is required to 
distinguish low dimensional chaotic 
data (which may have colored noise contamination) from a pure 
noisy time series.
The standard technique used to make such
distinctions is  surrogate data analysis with a 
discriminating statistic [20-22].
Here, a number of surrogate data are generated which have practically
the same power spectrum and/or the distribution of values as the
real data.  Then the real data and the surrogates are
subjected (in principle) to the same analysis to identify
if the real data is distinct from the surrogates. The null hypothesis is 
rejected if the discriminating statistic is different for the data and the 
surrogates. The most important measure used for the statistic is $D_2$, but the 
conventional method of calculating $D_2$ imposes major difficulties in 
implementing this standard procedure. As discussed by many authors [23-25], 
the standard correlation dimension analysis is
subjective since the scaling region used to compute $D_2$
has to be identified in a subjective manner. For some data, especially
those with inherent noise, different choices for the 
scaling region can significantly change the estimate. 
This subjectivity 
does not allow one to be certain that exactly the  same criteria have
been used to analyze both the real and surrogate data, which
is crucial to test the null hypothesis.

Several techniques have been proposed in 
the literature to overcome these difficulties. Judd [24, 26] has 
introduced an 
alternate method for the calculation of $D_2$ from the distribution of 
inter point distances in an attractor. This method avoids the 
problems associated with the scaling region, and among its many advantages  
is that it is 
particularly effective for attractors with multiple scaling. An 
extrapolation method has been proposed by Sprott and Rowlands[27]
to obtain a functional fit for
$D_{2}$ with $R$ as a convergence parameter. While the above methods
have their advantages, it is still useful to develop modifications
of the standard GP algorithm to address these issues, primarily
due to its wide spread applicability. 
Several attempts to improve the GP algorithm have been proposed and 
applied to some specific cases [28-34].  
These include applying a maximum likelihood 
estimator to the slopes taken at discrete points[29] and smoothing the 
$C_{M}(R)$ using a Gaussian kernel[30].
While such techniques have given robust results for particular systems,
like the human $\alpha$ rhythm [32],
they still suffer from either the subjectivity of choosing a 
proper scaling region, or are computationally complex.

This motivates us to propose and implement a modified GP algorithm which can
be effectively used for the analysis of noisy time series, especially for 
the surrogate analysis of  data involving colored noise.
The main modification in this approach is to fix  
the scaling region algorithmically, so that 
for a given finite time
series,  $D_2 (M)$ and the saturated correlation
dimension $D_2^{sat}$ can be computed in a ``non-subjective manner''.  
This ensures
that exactly the same procedure is used on the real
and the surrogate data.

In the next section,  the modified scheme is described while
in \S 3,  the scheme is tested using time series generated from a
number of standard analytical low dimensional chaotic systems and
pure colored noise.  It is confirmed that the computed 
$D_2 (M)$ and $D_2^{sat}$ are consistent with theoretical values. 
The variation of $D_2 (M)$ with the number of data points used in
the analysis is studied, and the expected increase in $D_2 (M)$ values with
addition of white and colored noise is verified. 
In \S 4, surrogate data analysis is performed for one
of these standard systems, with and without additional colored noise,
to ascertain under what conditions  the scheme can distinguish chaotic
from noisy data.
Quantifiers are introduced to measure the difference in 
$D_2$ values that can serve as benchmarking indices. As examples of real world 
applications,  data from two
different physical systems are analyzed in detail in \S 5 and  in \S 6,  
the main results of this work are summarized and discussed.

\section{Modified Algorithm}

The GP algorithm uses the delay embedding technique for the calculation of 
$D_2$. It creates an artificial space 
of dimension $M$ with delay vectors constructed by splitting a 
discretely sampled scalar time series $s(t_i)$ with 
\begin{equation}
   \vec{x_i} = [s(t_i), s(t_i+\tau), . ..., s(t_i+(M-1)\tau)]
   \label{eqn1}
\end{equation}
Here the delay time $\tau$ is chosen suitably such that the vectors are not 
correlated. The relative number of points
within a distance R from a particular ($i^{th}$) data point is given by
\begin{equation}
   p_{i}(R) = \lim_{N_v \rightarrow \infty} {1\over N_v} \sum_{j=1,j 
\neq i}^{N_v} H(R-|\vec{x_{i}}-\vec{x_{j}}|)
   \label{eqn2}
\end{equation}
where $N_v$ is the total number of reconstructed vectors and $H$ is the 
Heaviside step function.  Averaging this quantity over  randomly 
selected centers $N_{c}$ gives the correlation function
\begin{equation}
   C_{M}(R) =  {1\over N_{c}} \sum_{i}^{N_{c}} p_{i}(R)
   \label{eqn3}
\end{equation}
The correlation dimension $D_{2}(M)$ is then defined to be the scaling index 
of the variation of $C_{M}(R)$ with $R$ as $R \rightarrow 0$. That is, 
\begin{equation}
    D_2 \equiv \lim_{R \rightarrow 0} \frac{d\hbox {log} C_M (R)}{d\hbox {log} 
(R)}
    \label{eqn4}
\end{equation}
For a finite data stream of length $N$,  only a finite number of vectors
$N_v = N - (M-1)\tau$ can be constructed and hence the correlation 
dimension
computed will be a good approximation only if $N$ is large.  In practice,  
there are two other complications that hinder the 
accurate computation of $D_{2}(M)$.  First,  for small values of $R$, 
$p_{i} (R)$ would be small and hence will be 
affected
by noise due to counting statistics. 
Second,  for large values of $R$, 
a significant fraction of the  $M$-spheres used in the computation will
typically go beyond the attractor region.  This ``edge effect'' leads to
under estimation of $C_{M}(R)$ for large $R$ and finally 
causes $C_{M}(R)$ to saturate to unity.  To avoid these effects in 
practice, 
a proper linear part in the $\hbox {log} C_{M}(R)$ versus $\hbox {log} R$ 
plot is identified
which is called the ``scaling region'' and its slope is taken to be $D_{2}$.
However, such an exercise is subjective, being specific to data, especially for 
higher values of M.

In our scheme,  the original data set, $s_{t_i}$, is first transformed 
to a uniform deviate, $s_{u}(t_{i})$. 
Note that 
$s_{u}(t_{i})$ ranges from $0$ to $1$, which makes the volume of the 
embedding 
space unity.  In order to take into account the edge effects 
correctly, it is 
convenient to redefine  $p(R)$ as the  
number of data points within a M-cube (instead of M-sphere) of length 
$R$ around a data point. 
This is equivalent to replacing the Euclidean norm by the 
maximum norm. Operationally this is done by choosing randomly $N_{c}$ 
data 
points as centers of M-cubes of length $R$.  Of these $N_{c}$ 
M-cubes, only those
 are considered which are within the bounding box of the embedded data.  
Finally the 
correlation sum $C_{M} (R)$
is obtained by averaging the number of data points within the 
M-cubes. 
The imposition that a M-cube has to be within the embedding space 
ensures that there are no edge effects due to limited data points. 
However,  this also means that for large values of $R$,  only a 
small fraction of 
the original $N_{c}$ M-cubes are taken into consideration.  Hence a 
maximum 
value of $R$,  $R_{max}$,  is fixed such that for 
all $R < R_{max}$ the number of M-cubes which satisfy the above 
criterion,  is
at least one-hundredth of the total number of vectors,  i. e.  $N_v/100$. 
To avoid the region dominated by counting statistics  
only results from $R > R_{min}$ are taken 
into consideration,  where $N_v C (R) > 10$,  which ensures that on
the average at least ten data points are considered per center. 
This makes sure that the region $R_{min} < R < R_{max}$ is not affected
by either ``edge effects'' or counting statistics. Although the
criteria used to compute $R_{min}$ (i.e.  $N_v C (R) \approx 10$) 
and $R_{max}$ (i.e. number of M-cubes $\approx N_v/100$) may not be
optimal for every kind of system, they do provide good estimates
for all the systems  we have studied in the next section.
Moreover, from a surrogate analysis point of view (see \S 4), fixing
these criteria  a priori, ensures that the same conditions 
are imposed on the algorithm for estimating $D_2$ of the data 
and the corresponding surrogates.  $C_{M}(R)$ 
is computed for several different values of $R$ between $R_{max}$ and 
$R_{min}$,  the logarithmic slope at each point is calculated and the 
average is taken to be $D_{2}(M)$.  The error on $D_{2}(M)$ is estimated 
to be the mean standard deviation over this average. This error
is an estimate of how well the region used by
the scheme, $R_{min} < R < R_{max}$, can be represented 
by  a linear scaling region. A large error signifies that,
those values of $R$ for which $C_{M}(R)$ is not affected by
counting statistics and edge effects, do not represent a single
scaling region. 
It should be noted that 
there 
often exists a critical embedding dimension $M_{cr}$ for which $R_{min} 
\approx R_{max}$ and no 
significant results can then be obtained for $M > M_{cr}$.  Thus our algorithm 
fixes an upper limit on $M$ up to which calculations are to be repeated. 
For practical implementation of the above scheme, 
it is sufficient to choose $N_c$ as $0. 1 N_v$.  
The delay time  $\tau$ is chosen to be the value 
where
the auto-correlation function drops by $1/e$.  With these values,  
$D_2(M)$
for $M = 1$ to $M = M_{cr}$ is computed for a given data stream and
a chi-square fitting is undertaken using a simple analytical function
\begin{eqnarray}
f(M)\; &  = & \;\Big {(}{D_2^{sat} - 1 \over M_d -1}\Big {)} (M-1) +1 
\;\;\;\; \hbox  {for} \;\;\;\;  M < M_d \nonumber \\
                   &  = & \; D_2^{sat} \;\;\;\;  \hbox  {for} \;\;\;\;  M 
\geq M_d
\label{fit}
\end{eqnarray} 
The best fit value of $D_2^{sat}$ (obtained by minimizing $\chi^2$)
is taken to be the saturated correlation
dimension with errors corresponding to $\Delta \chi^2 = 1$.  Considering
the uncertainties in the computation and statistics of the errors in
$D_2 (M)$,  a more sophisticated fitting procedure is perhaps not 
warranted. A best fit value of $D_2^{sat} \approx M_{cr}$ implies
that no saturation of $D_2 (M)$ was detected.

In summary,  the algorithmic scheme first converts a data stream to
a uniform deviate.  Next, the autocorrelation function is evaluated to
estimate the time delay $\tau$.  For each $M$,  the $C_M(R)$ are evaluated
using $N_c = 0. 1 N_v$ randomly chosen centers.  The limits 
$R_{min}$ and $R_{max}$ are estimated and $D_2 (M)$ is computed for
the region from $R_{min}$ to $R_{max}$ .  The process is repeated for 
consecutive values of
$M$ till $R_{max} \approx R_{min}$.  The resultant $D_2 (M)$ curve is
fitted using function (\ref{fit}) which returns the saturated 
correlation
dimension $D_2^{sat}$ with an error estimate. A numerical code
which implements the scheme is available at 
http://www.iucaa.ernet.in/~rmisra/NLD.

\section{Synthetic Data Analysis}

To illustrate the applicability of the scheme, it is used to analyse  
synthetic time series
generated from six well known  low dimensional chaotic 
systems. 
Figure 1 shows the results of the analysis for a $30000$ 
points long data stream 
generated from the Rossler system.  The solid lines in the $C_M(R)$ 
versus
$R$ curves mark the region between $R_{min}$ and $R_{max}$ that the
scheme uses to compute $D_2 (M)$.  In the standard scheme these regions
are usually selected subjectively by the analyst.  As expected,  
the $D_2 (M)$ curve is significantly different from that of white
noise, where $D_2 (M) = M$. Hence  curve is fitted using the 
function defined in (\ref{fit}).  The best fit function is shown as 
a dashed line
along with $D_2 (M)$.   The scheme returns $D_2^{sat} = 1. 97 \pm 0. 17$
which may be compared with the value $1. 99 \pm 0. 08$ reported in the 
literature[27].  The exercise
is repeated for five other standard systems and the resultant 
$D_2(M)$ curves are shown in Figure 2.  The $D_2^{sat}$ values for
all the six systems computed by our scheme are shown  in Table 1,  
along with the values taken from [27] for comparison in the last column. 
The number of points used in all cases are $30000$. 

\begin{figure}
\begin{center}
\includegraphics*[width=14cm]{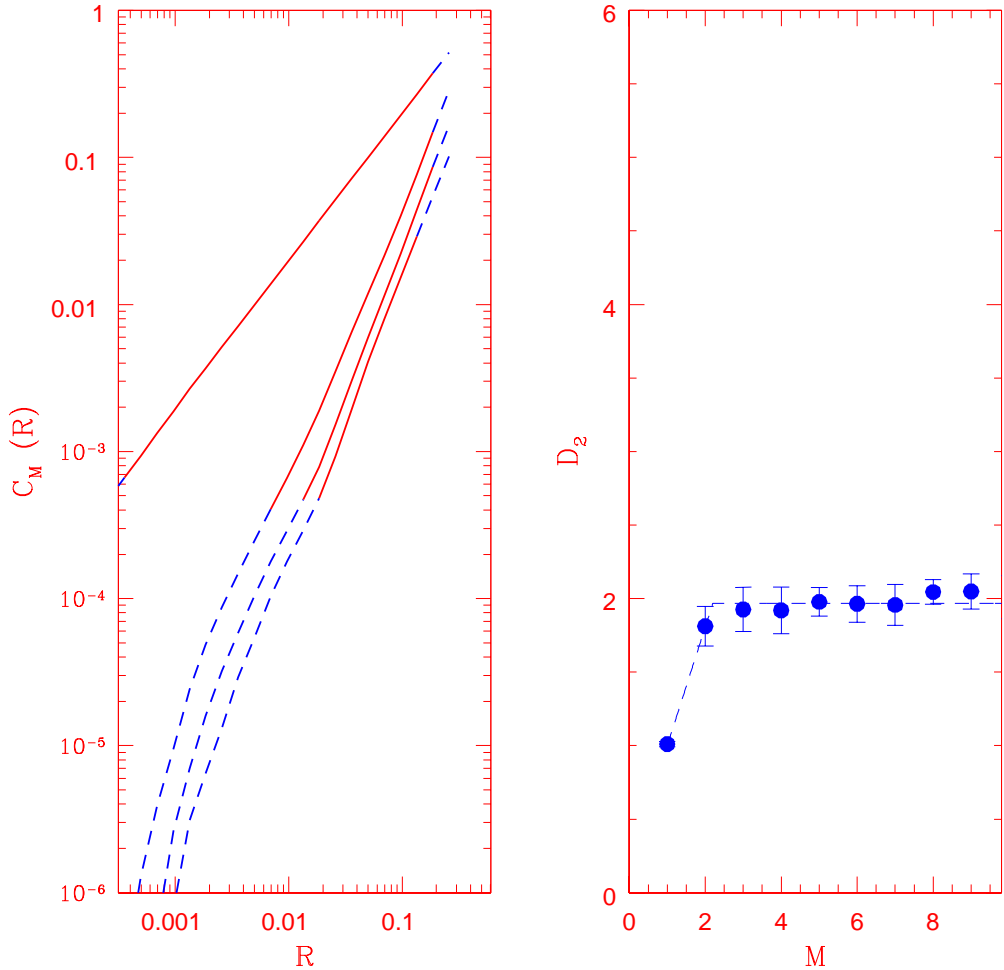}
\end{center}
\
\caption{ The results of the analysis for $30000$ points from the 
Rossler system. Left Panel: The $C_M (R)$ curves for $M = 1,2,5$ and $9$ (from
left to right). 
The solid lines
indicate the scaling region which the scheme uses to compute $D_2 (M)$. 
Right 
Panel: The $D_2 (M)$ values (filled circles). 
The dashed line is the best fit curve, 
giving
$D_2^{sat} = 1.97 \pm 0.17$.  }
\label{Fig.1label}
\end{figure}

\begin{table}
\caption{}

\begin{center}

\begin{tabular}{ccc} \hline
\emph{System} 	& Computed ${D_{2}^{sat}}^\ast$ 	& 
{${D_2^{sat}}^{\dagger}$} \\ \hline \hline
Rossler attractor & 			& 		\\
$a=b=0.2,c=0.78$ & $1.97 \pm 0.17$ & $1.99 \pm 0.08$ \\ 

& & \\

Lorenz attractor & 			& 		\\
$\sigma=10,r=28,b=8/3$ & $2.03 \pm 0.16$ & $2.05 \pm 0.1$ \\ 

& & \\

Ueda attractor & 			& 		\\
$k=0.05,A=7.5$ & $2.59 \pm 0.1$ & $2.67 \pm 0.13$ \\ \hline

& & \\

Henon map & 				& 		\\
$a=1.4,b=0.3$ & $1.23 \pm 0.1$ & $1.22 \pm 0.04$ \\ 

& & \\

Lozi map & 				& 		\\
$a=1.7,b=0.5$ & $1.41 \pm 0.1$ & $1.38 \pm 0.05$ \\ 

& & \\

Cat map 	& $2.00 \pm 0.01$ & $2.00 \pm 0.06$ \\ \hline

\end{tabular}

\end{center}
$\ast $ {$D_2^{sat}$ values obtained by using the scheme prescribed in
this work. For all data sets the number of points used are $30000$}

$\dagger $ {$D_2^{sat}$ taken from [27] }

\end{table}

For a  randomly generated data set (i.e. white noise), the  $D_2(M)$ curve 
goes as $D_2 (M) = M$ which is also shown in Figure 2. However,for colored
random noise characterized by a power-law spectrum, the $D_2 (M)$ values 
saturate with $M$.  
The prescribed scheme should also reproduce effectively  $D_2^{sat}$
for such non-chaotic systems.  This is demonstrated in Figure 3,  where
the $D_2 (M)$ curves are plotted for colored noise corresponding to 
three
different values of $\alpha$.  In all three cases,  the computed values agree 
with the theoretically expected values. 

\begin{figure}
\begin{center}
\includegraphics*[width=14cm]{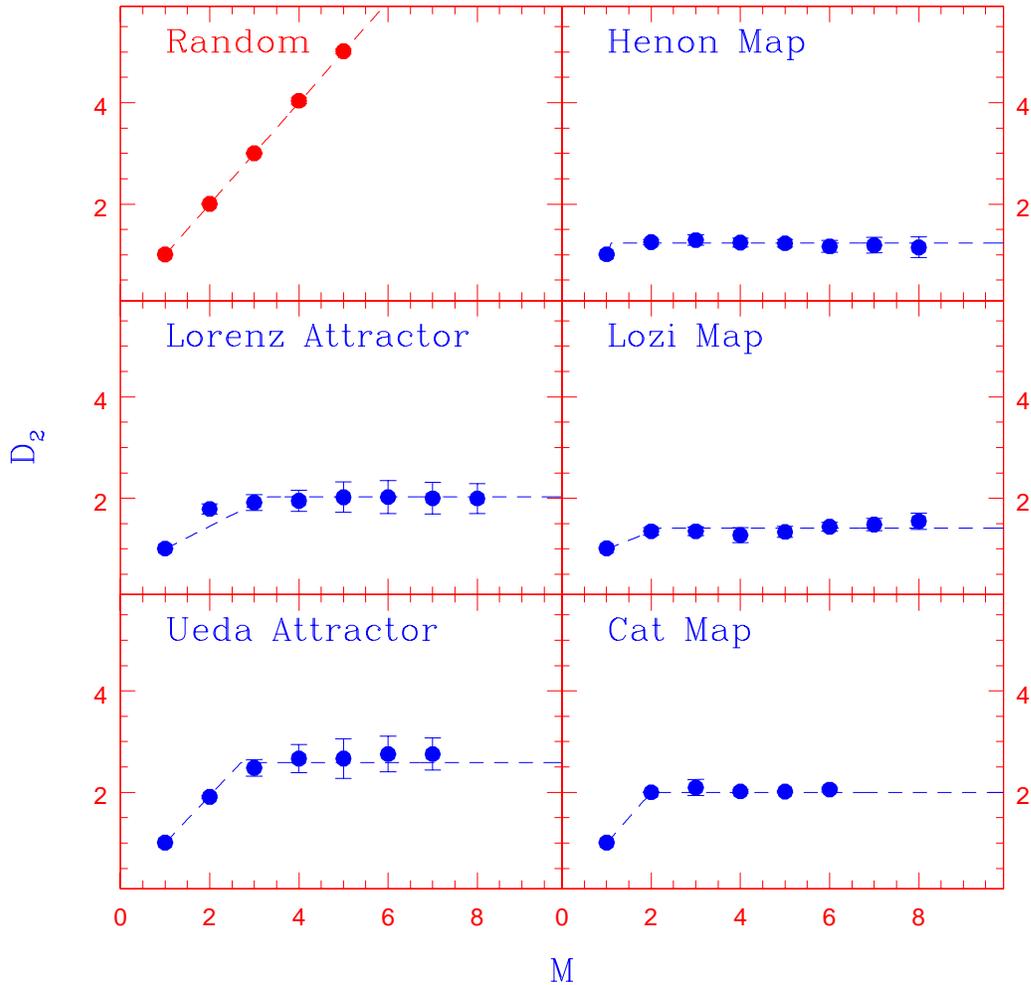}
\end{center}
\
\caption{ The $D_2 (M)$ values (filled circles) 
for random data (white noise) and
five standard low dimensional chaotic systems each having $30000$ 
points. 
The dashed lines represent
the best fit curves giving $D_2^{sat}$ values tabulated in 
Table 1. }
\label{Fig.2label}
\end{figure}

The results of the analysis are expected to depend
on the number of points in the data stream. As examples,  
we study the dependence of the computed saturated dimension $D_2^{sat}$ 
on the number of data points available for the Rossler,  
Ueda and Henon systems. 
For the flows  Rossler and Ueda,  the results depend also on the
total time $T$  to which the governing differential equations have been
evolved.  If $T$ is small,  then the system would not have traced
out its entire attractor region and hence incorrect value of
$D_2 (M)$ would be computed even if the number of points
available is large.  In Figure 4(a),  the variation of
$D_2^{sat}$ with N are plotted.  Here the total time is held
constant at  $T = 300$ and $1500$ for the Rossler and Ueda systems 
respectively.  The 
data is then sampled at different time intervals $\delta T$,  
to obtain different values of the 
number of available points i.e.  $ N = T/\delta T$.  Figure 4(b)
shows the dependence of $D_2^{sat}$ on the total time $T$,  when 
the number of data points is held constant at $N = 10000$.  As can be
seen in the figure,  for the Rossler system,  the technique 
computes reasonable values of $D_2^{sat}$ even when $N \sim 1000$,  provided
the total time $T \geq 300$.  However,  these values depend on 
the system and the type of attractor,  since for the Ueda system, 
reasonable values are obtained only when $N \geq 20000$ and
the total time $T \geq 800$.  This is because of the difference in the 
inherent Poincare times involved. For maps and colored noise,  there
is no inherent time-scale in the system and hence the only relevant
parameter is the number of data points.  The variation of $D_2^{sat}$ 
for the Henon map is also shown in Figure 4(a).

\begin{figure}
\begin{center}
\includegraphics*[width=14cm]{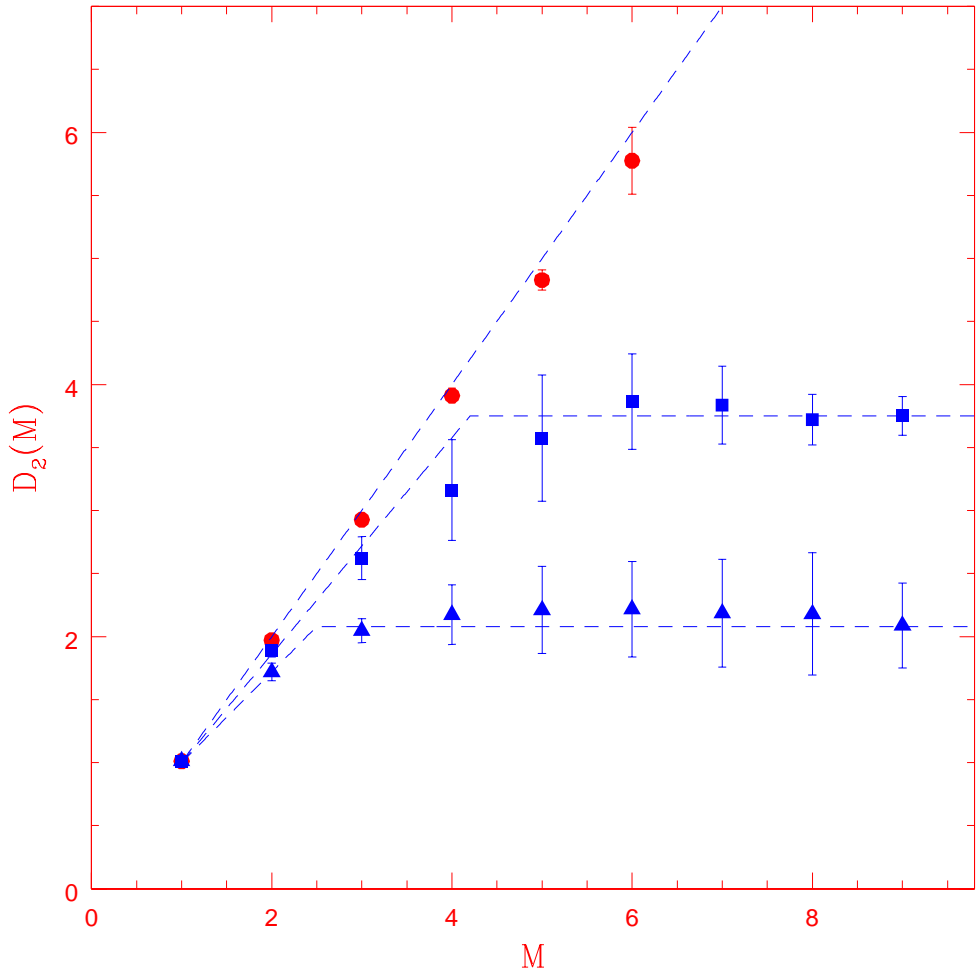}
\end{center}
\
\caption{ The $D_2 (M)$ values for colored noise data, each having $30000$ 
points, corresponding to spectral index $\alpha = 1$ (circles), 
$1.5$ (squares) and $2$ (triangles). The dashed lines are the 
best fit curves. For $\alpha = 1$ the curve is consistent with white noise 
while for
$\alpha = 1.5$ and $2.0$, 
$D_2^{sat} = 3.75 \pm 0.25 $ and $ 2.08 \pm 0.16$ respectively. The 
theoretical value is $D_2^{sat} = 2/(\alpha-1)$, which gives $\infty$, 
$4$ and $2$ for
$\alpha = 1.0$, $1.5$ and $2.0$ respectively  }
\label{Fig.3label}
\end{figure}

\begin{figure}
\begin{center}
\includegraphics*[width=14cm]{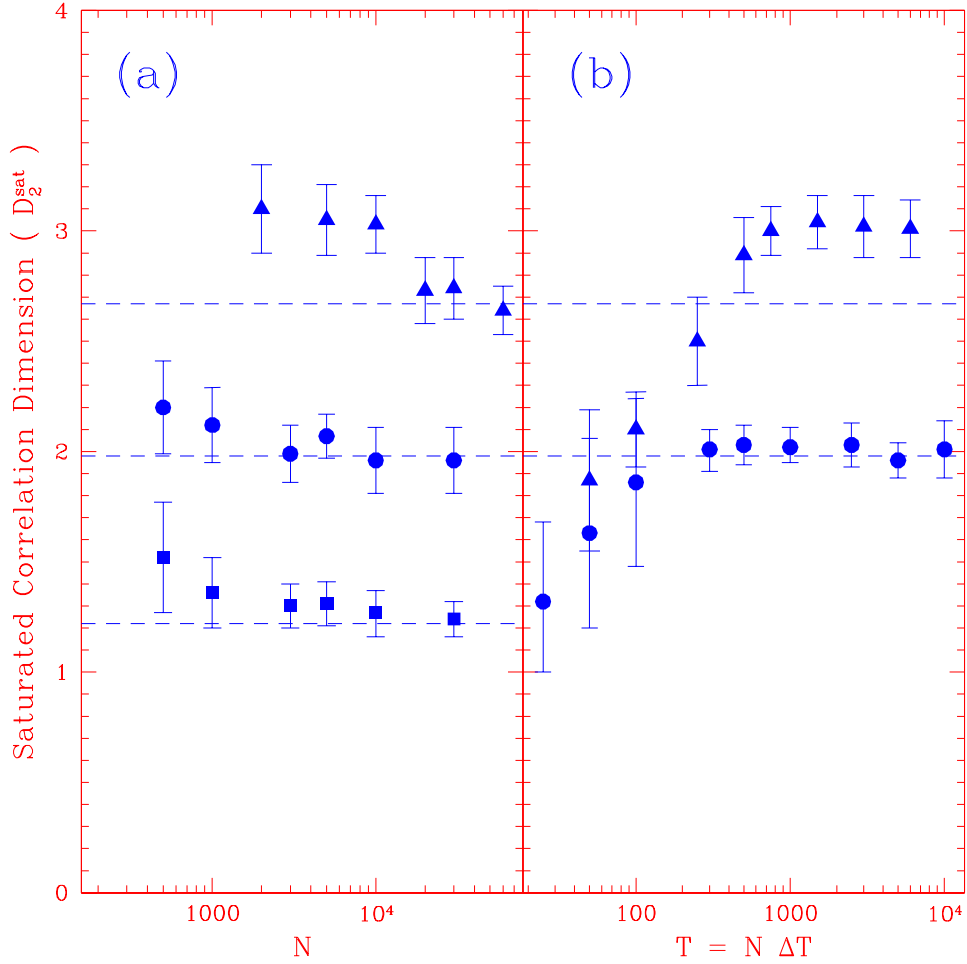}
\end{center}
\
\caption{ (a) The saturated correlation dimension $D_2^{sat}$ versus
number of data points $N$ for Ueda (triangles), Rossler (circles) and
Henon (squares) system. The total time evolved is $T = N \delta T = 300$
and $1500$ for the Rossler and Ueda systems respectively. (b)
The saturated correlation dimension $D_2^{sat}$ versus the total
time $T$ for the Ueda (triangles) and  Rossler (circles) system. The total
number of points is fixed at $N = 10000$. 
The dashed lines indicate the standard $D_2$ values.  }
\label{Fig.4label}
\end{figure}

\begin{figure}
\begin{center}
\includegraphics*[width=14cm]{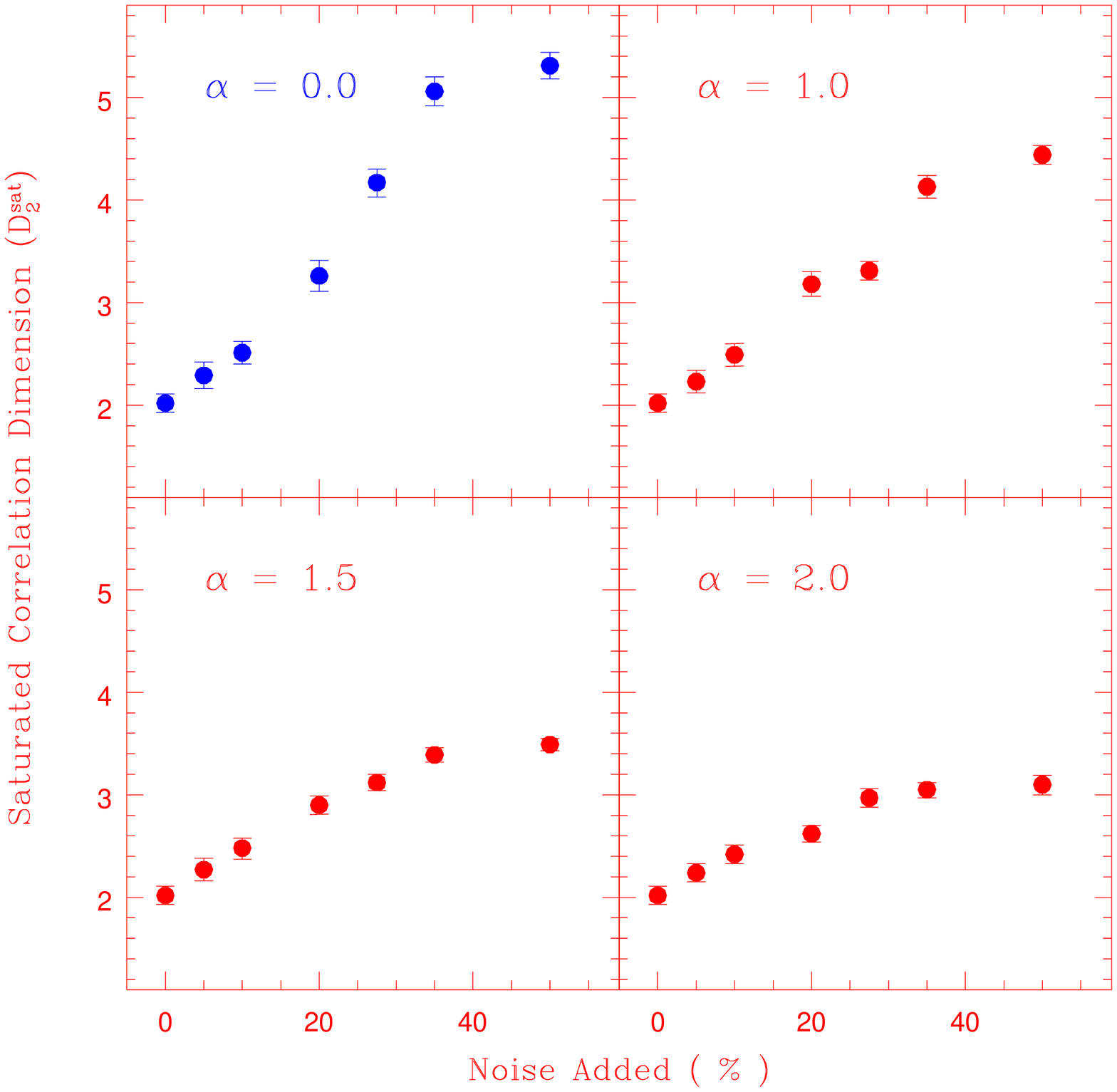}
\end{center}
\
\caption{ The saturated correlation dimension $D_2^{sat}$ for
the Rossler system with different percentage of noise added. The four
figures correspond to the addition of colored noise with different
power spectrum index $\alpha$. For white noise $\alpha = 0$ and
red noise has $\alpha = 2$. The number of points in the data stream is
$10000$.  }
\label{Fig.5label}
\end{figure}

The addition of noise to the data from a chaotic system is expected to
increase the value of $D_2 (M)$.  This is illustrated in Figure.5
where $D_2^{sat}$ has been plotted for the Rossler system with different
percentages of white and colored noise added.  Since colored noise has
intrinsically low
values of $D_2 (M)$,  the increase in $D_2^{sat}$ is less for addition
of colored noise as compared to white noise.  
The addition of even 50\% red noise ($\alpha = 2. 0$) 
increases $D_2^{sat}$ only up to $\sim 3$.  This emphasizes the need for
surrogate data analysis to differentiate between chaotic 
systems contaminated with noise, from purely stochastic ones.

\begin{figure}
\begin{center}
\includegraphics*[width=14cm]{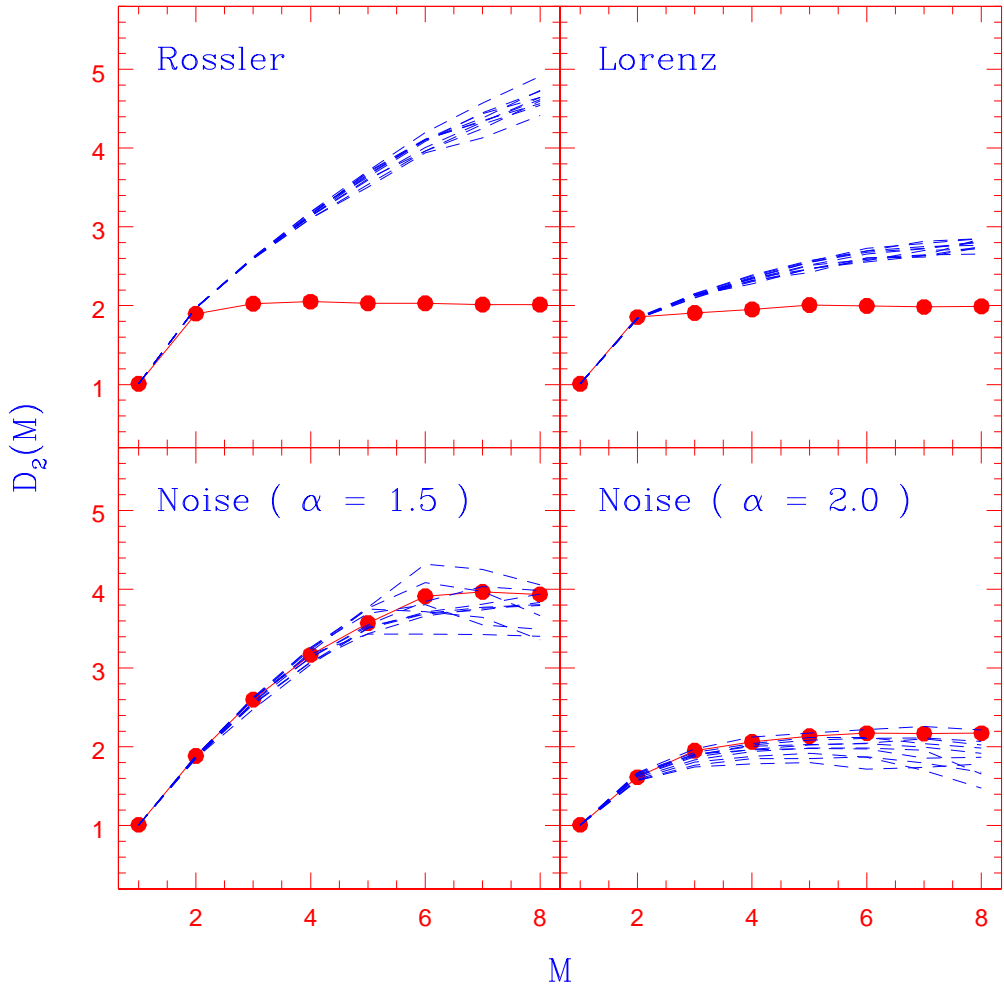}
\end{center}
\
\caption{ Correlation dimension $D_2$ versus $M$, for two chaotic
systems ( Rossler and Lorenz) and for colored noise with $\alpha = 1.5$ and
$2.0$ ( filled circles connected by solid lines). The corresponding 
$D_2 (M)$ for ten surrogates are shown by 
dashed lines. The error bars are omitted for clarity. 
The normalised mean sigma deviation, $nmsd$ (see text) between the
real and surrogate data for the chaotic systems Rossler and Lorenz are
$ nmsd = 36.1 $ and $10.2$ respectively, while for the colored noise 
$ nmsd = 0.68$ and $2.00$ for $\alpha = 1.5$ and $2.0$ respectively.} 
\label{Fig.6label}
\end{figure}

\section{Surrogate Data Analysis}

Surrogate data analysis is perhaps the first important
analysis that needs to be undertaken on a time series
data to detect the presence of non-trivial structures. 
The basic idea  is to formulate a null hypothesis 
that the data has been created by a stationary linear stochastic process, and 
then to attempt to reject this hypothesis by comparing results for the 
data with appropriate realizations of surrogate data. Ideally,
surrogate data sets should have
the same power spectrum and  distribution of values as the real data.
The method to generate surrogate data, namely 
Amplitude Adjusted Fourier Transform (AAFT), was originally proposed by
Theiler and co-workers [20]. But recently Schreiber and Schmitz [21, 22]
have proposed an iterative scheme, known as IAAFT, which is similar but
reported to be more consistent in representing the null hypothesis [35].
In this work, we apply this scheme to generate ten 
surrogate data sets for each analysis, using the TISEAN package [36,37].

\begin{figure}
\begin{center}
\includegraphics*[width=14cm]{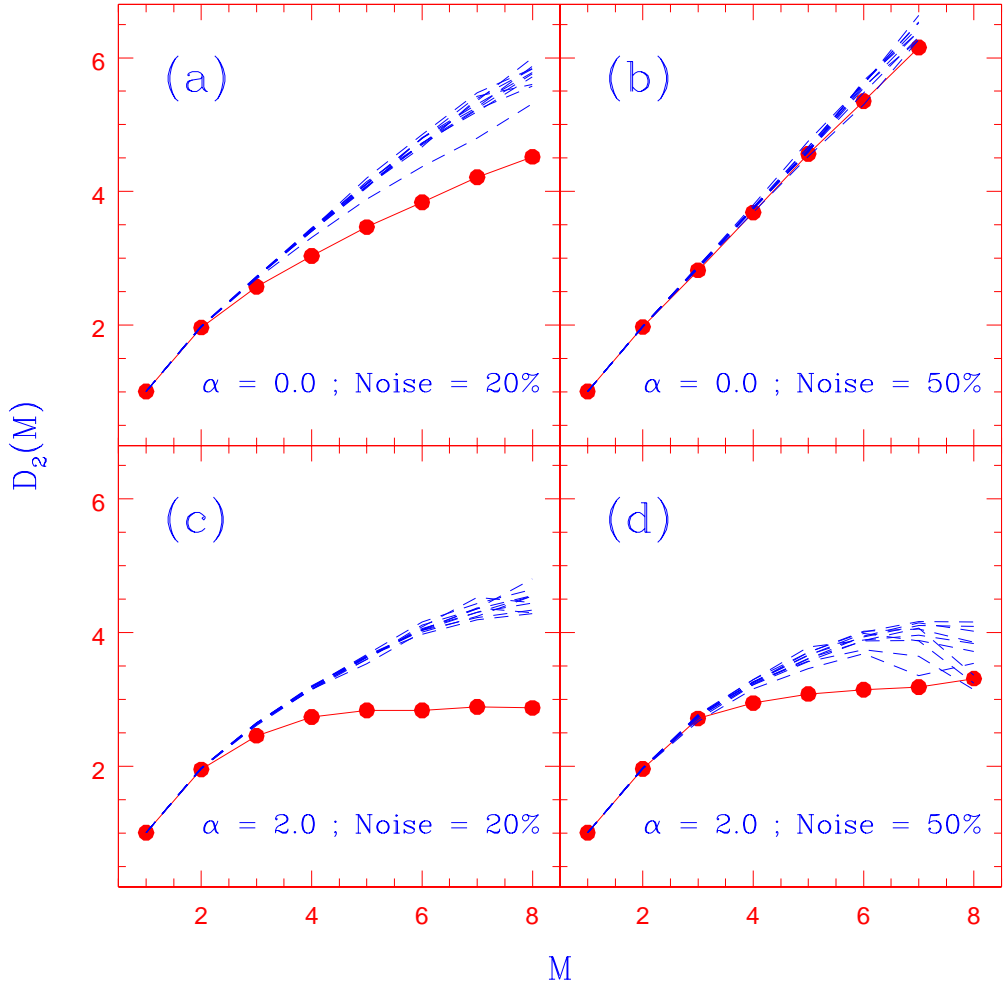}
\end{center}
\
\caption{ The effect of addition of white ($\alpha = 0$) and 
red noise ($\alpha = 2.0$) on
surrogate analysis of data from the Rossler system. The $D_2 (M)$ values
for the data are represented by filled circles and connected by solid lines,
while for the corresponding ten surrogates, the curves are represented  by dashed lines. 
The upper panel (a) and (b) are for
white noise contamination at 20\% and 50\% reprectively.
The lower panel (c) and (d) are for
red noise contamination at 20\% and 50\% reprectively.   }
\label{Fig.7label}
\end{figure}

\begin{figure}
\begin{center}
\includegraphics*[width=14cm]{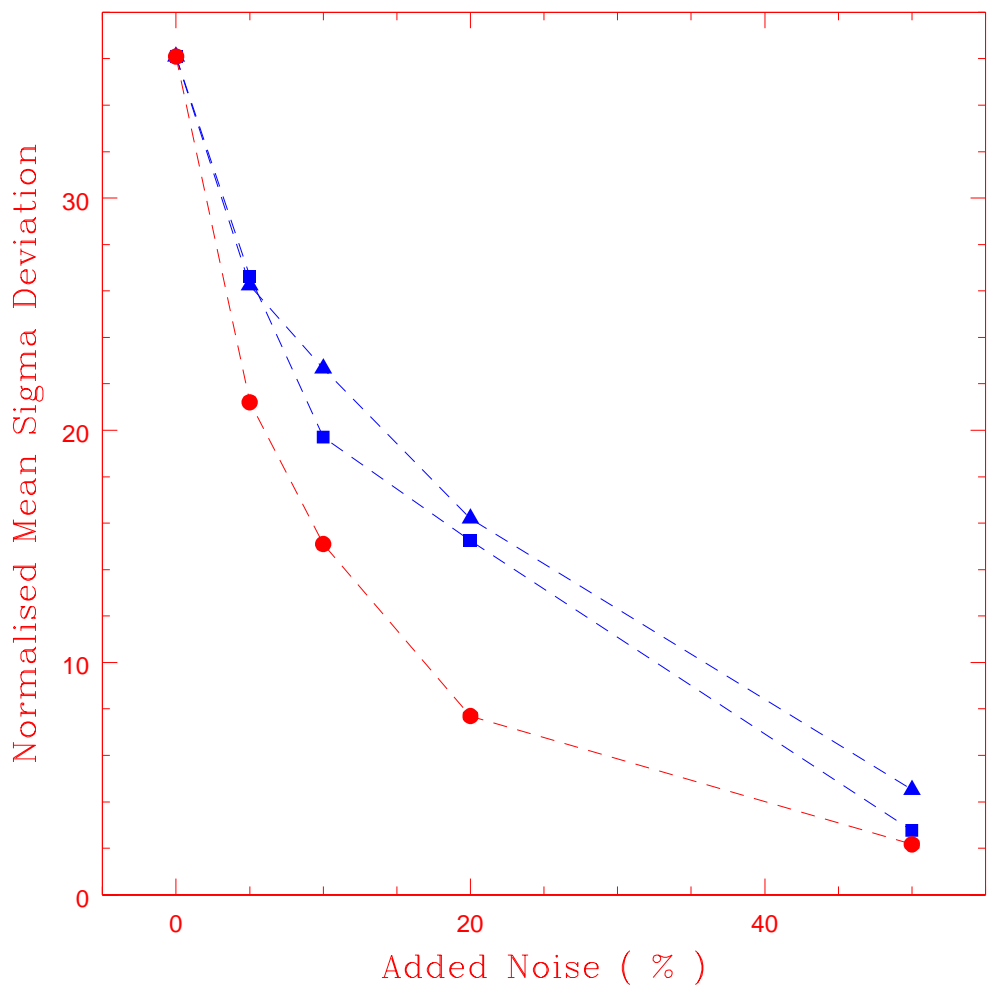}
\end{center}
\
\caption{ The normalised mean sigma deviation  $nmsd$ between data from  
the Rossler system and ten surrogates versus the percentage of noise
added to the data. Circles : white noise ($\alpha = 0$); Squares:
colored noise ($\alpha = 1.5$); Triangles: colored noise ($\alpha = 2.0$).
 }
\label{Fig.8label}
\end{figure}

In Figure 6,  the $D_2 (M)$ values for data from two chaotic
systems (Rossler and Lorenz) and two kinds of pure colored noise,  are shown along
with the corresponding surrogates. In all analysis of this section,
$10000$ data points are used.  As expected the results for
the chaotic systems show clear deviation from their surrogates,  while
for pure colored noise,  the results are similar to the surrogates 
and hence the null 
hypothesis cannot be rejected.  The addition of noise
to the chaotic system is found to decrease the difference between
the $D_2 (M)$ of the data and the surrogates.  This is shown in
Figure 7,  where $D_2 (M)$ for data and surrogates are
compared for the Rossler system having different percentages of
red colored and white noise added to the time series.  Visual inspection of
Figure 7 (upper panel),  reveals that when white noise is added to the system, 
$D_2 (M)$ for both the data and the surrogates increases.  There is
noticeable difference between the data and surrogates for a contamination
level of 20\%,  but for a larger level of 50\%,  the data and the surrogates
are no longer distinguishable.  Red noise contamination is more
interesting (Figure 7, lower panel), since
for pure red noise (i. e.  $\alpha = 2. 0$) the saturated
correlation dimension $D_2^{sat} =  2$, which is roughly the same as that 
for the Rossler system.  Thus the $D_2 (M)$ value for surrogates
\emph{decreases}  as the 
percentage of contamination increases,  while 
the $D_2 (M)$ values for the data increase.  Nevertheless,  even
for a noise added level of 50\%, the data and the surrogates
are still distinguishable,  in contrast to the case when white noise
was added to the system. This result has also been verified for other 
values of $\alpha$.

The above mentioned differences may appear highly qualitative and hence a 
quantification is attempted by defining a normalised mean sigma 
deviation, $nmsd$. 
For this the average of $D_2^{surr} (M)$, denoted here as
$< D_2^{surr} (M) >$, is estimated using a number of 
realizations of the surrogate data. Then
\begin{equation}
nmsd^2   = \frac{1}{M_{max} -1} \sum_{M = 2}^{M_{max}} \Big {(}\frac{D_2 (M) - < D_2^{surr} (M) >}{\sigma^{surr}_{SD} (M) }\Big {)}^2 
\end{equation}
where $M_{max}$ is the maximum embedding dimension for which the analysis
is undertaken and $\sigma^{surr}_{SD} (M)$ is the standard
deviation of $D_2^{surr} (M)$.  The normalised mean sigma deviation, $nmsd =  36. 1$
and $10. 2$ for the Rossler and Lorenz systems respectively,  while 
 $nmsd =  0. 6$
and $2. 0$ for the two cases of colored noise (Figure 6).  Thus
a conservative upper limit of $nmsd = 3$ can be imposed,  such that
when  $nmsd$ is greater than $3$,  the data can be considered to be
distinguishable from its surrogates. 
The variation of $nmsd$ for different percentages of white and colored 
noise added to the Rossler system is shown in Figure 8. 
Note that for the same percentage contamination,  
$nmsd$ is larger for colored noise than for white noise.

\section{Experimental Data Analysis}

\begin{figure}
\begin{center}
\includegraphics*[width=14cm]{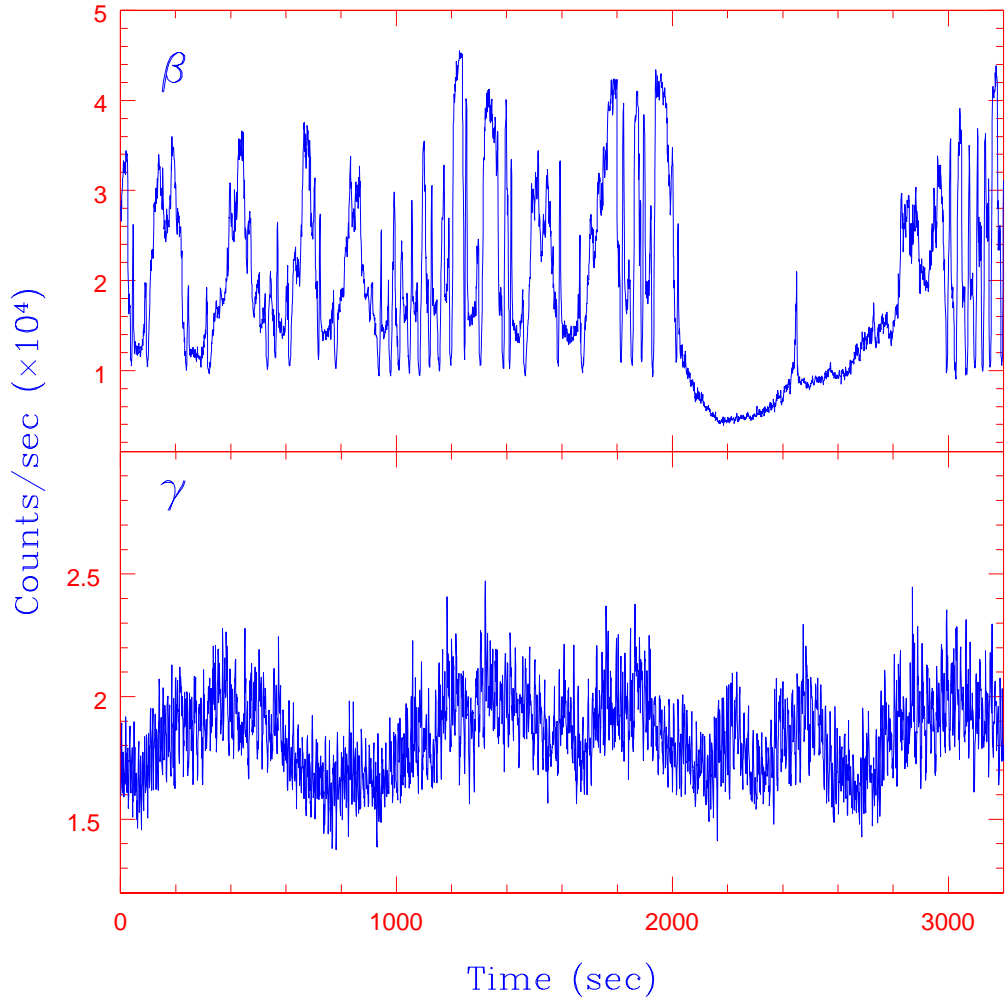}
\end{center}
\
\caption{The light curves from two temporal states, $\beta$ and $\gamma$ 
 of the black hole system GRS1915+105, see text for details.   }
\label{Fig.9label}
\end{figure}

\begin{figure}
\begin{center}
\includegraphics*[width=14cm]{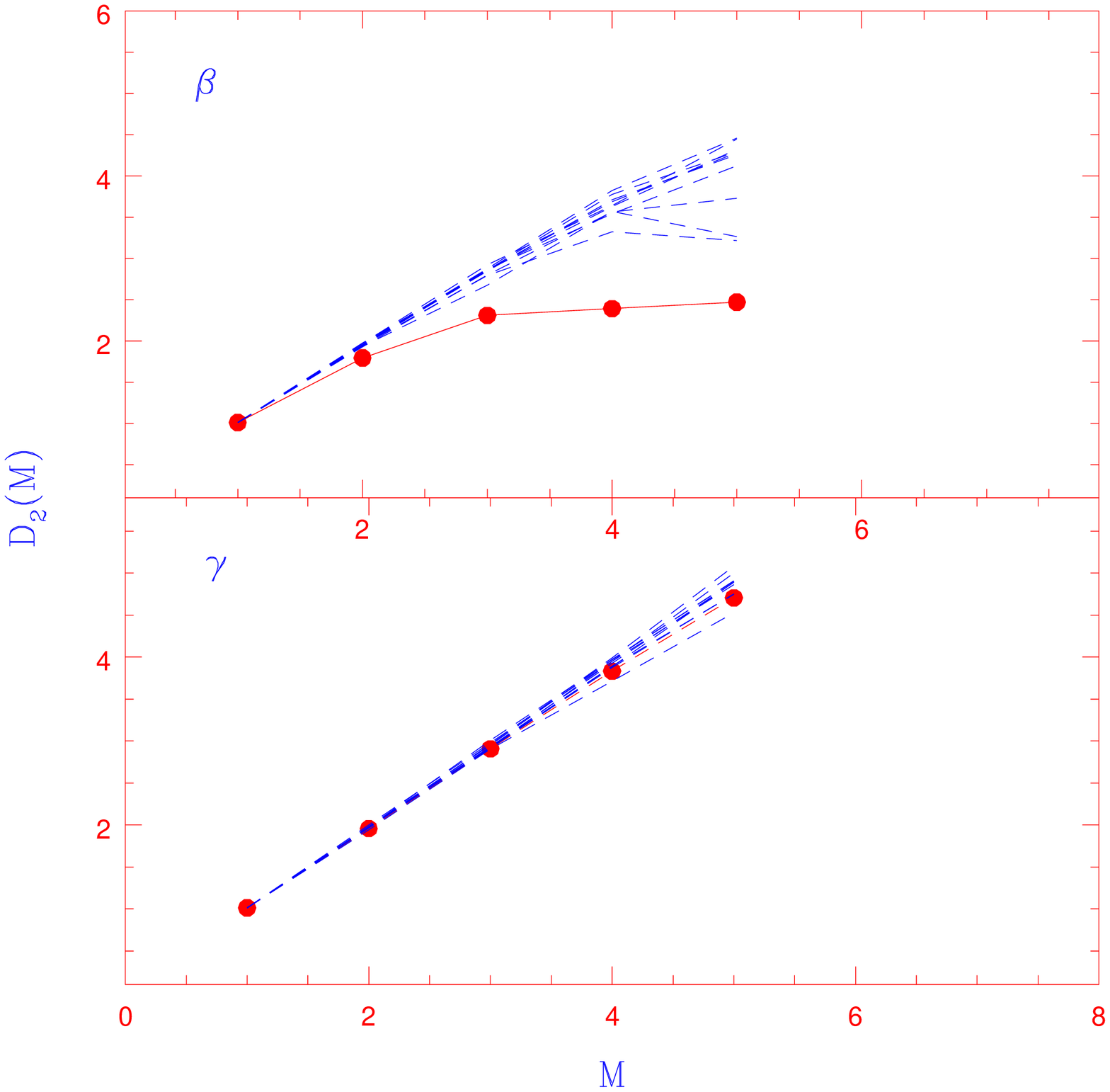}
\end{center}
\
\caption{The $D_2 (M)$ values for the data (filled circles connected by line) 
and the corresponding values for surrogates (dashed lines) for the 
states shown in Figure 9}
\label{Fig.10label}
\end{figure}

Finally, we consider two data streams each obtained from different real world 
experiments, one set from an astrophysical X-ray object and the other,  EEG 
data of the human brain. For both
these cases,  the continuous data stream has $ N < 5000$ points and
is expected to have noise contamination of unknown type and percentage. 
The X-ray data is taken from GRS 1915+105, which is a highly variable 
black hole system. Its temporal behavior has been classified into 12 different 
states and it shows signatures of low
dimensional chaotic behavior in some of these states [38]. Here we 
choose representative data sets from two different spectral classes, namely, 
the $\beta$ state and the $\gamma$ state, both
consisting of 3200 data points. The data has been extracted with a resolution 
of one second to avoid the effect of Poisson noise. The X-ray light curves are  shown 
in Figure 9, while Figure 10 shows the $D_{2}(M)$ curves for the data and the 
surrogates in both cases, obtained by applying our scheme. Here the 
$nmsd$ is found to 
be 7.02 and 0.89 for $\beta$ and $\gamma$ respectively, indicating that the 
null hypothesis can be rejected for the $\beta$ state.  

There are indications of 
nonlinear signature in the dynamical properties of the human brain's 
electrical activity, particularly during an epileptic seizure. 
Hence we choose two EEG data sets of the human brain, one during seizure period 
and the other for healthy state; both  consist of 4098 data points. 
The data has been studied earlier and further details regarding its 
analysis can be found in Andrzejak et. al [39]. The EEG time series for both 
cases are shown in Figure 11, and 
the results of our analysis are shown in Figure 12. The curves 
are clearly different for the data and the surrogates for the seizure signal 
with $ nmsd = 20.1$, while the healthy signal behaves as noise with 
$ nmsd = 0.64$, consistent with the earlier analysis in [39]. 
 
\begin{figure}
\begin{center}
\includegraphics*[width=14cm]{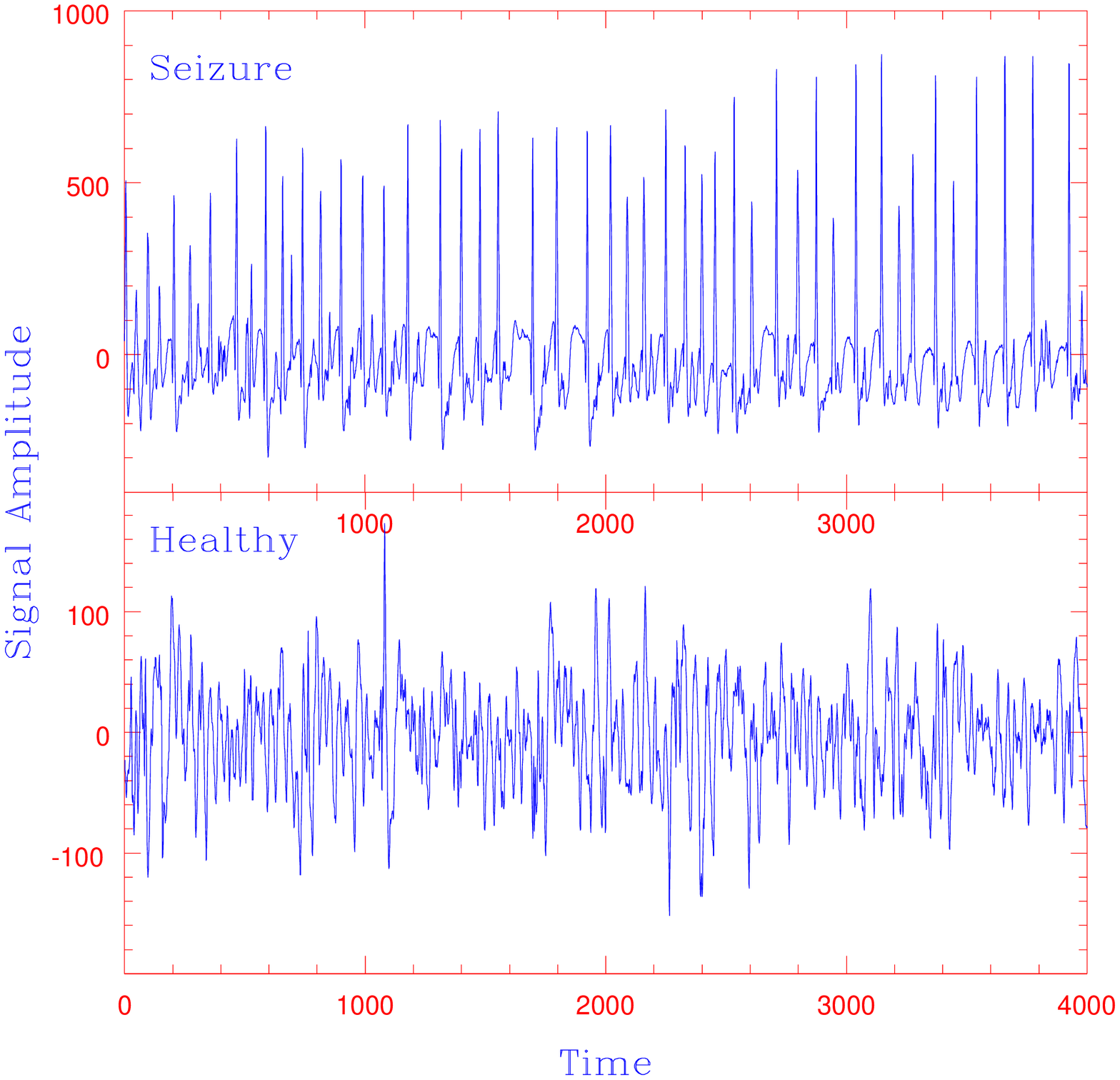}
\end{center}

\caption{The EEG time series for a seizure period and the healthy state}
\label{Fig.11label}
\end{figure}

\begin{figure}
\begin{center}
\includegraphics*[width=14cm]{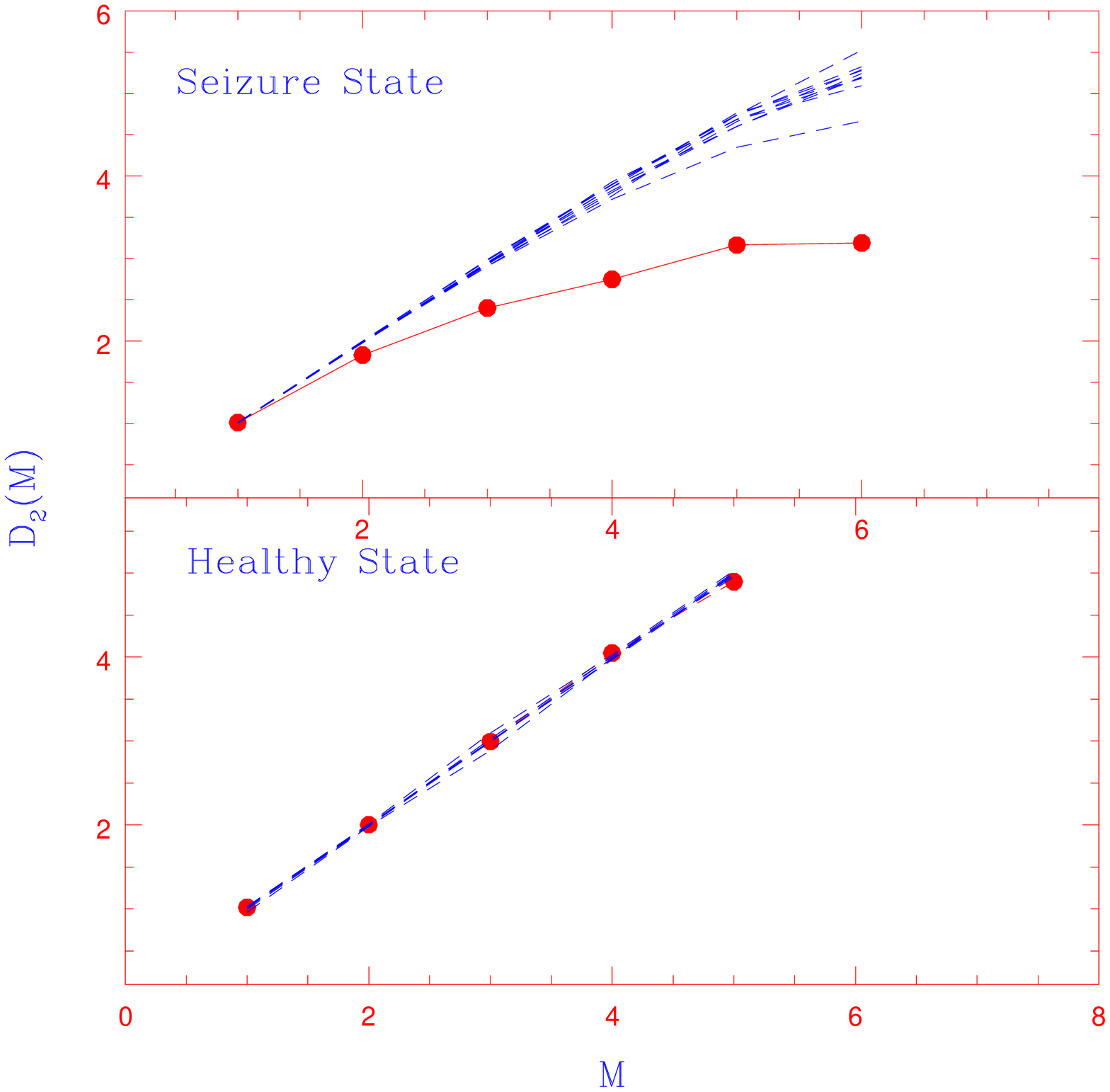}
\end{center}

\caption{The $D_2 (M)$ values for the data(filled circles) and the corresponding 
surrogates for the two EEG time series shown in Figure 11}
\label{Fig.12label}
\end{figure}

\section{Discussion}

In this paper we have implemented a modification of the conventional 
GP algorithm and
calculated the correlation dimension in a non-subjective manner.  
The method is most suitable for surrogate data analysis where it is 
imperative that the same conditions are maintained in the algorithm for the 
data and the surrogates. 
Moreover, it can be applied to any arbitrary time series with a few thousand 
data points and provides an error estimate on the value of $D_2^{sat}$ obtained.

The scheme is
tested for standard low dimensional chaotic systems and
for pure colored noise, and it is found that the computed
$D_2 (M)$ are close to the standard values in all cases. 
As expected, the addition of noise to the data from chaotic systems
increases the correlation dimension $D_2 (M)$.  The scheme
can differentiate between the results from
noise contaminated data and corresponding surrogates,  when
the percentage of noise addition is low.  The level of noise
contamination up to which this differentiation can be made
depends on the type of  noise.  The effect of  
contamination by red noise (which intrinsically has a low
saturated correlation dimension) on surrogate
data analysis is less compared to the addition of white noise,  
i. e.,  for the same percentage
of noise addition,  data with colored noise is more distinguishable
from corresponding surrogates,  than data with white noise.  This
implies that in those practical situations where experimental
data have colored (or an unknown type of) noise combined
with the real signal, the present scheme can be used  effectively.  As
examples of application,  data sets from two scientific experiments are analyzed 
and the nature of their variability ascertained.

It is also important at this stage to highlight a few  possible
limitations of this scheme while applying it to specific data sets.  
The scheme assumes that the actual
scaling of $C(R)$ with $R$ is independent of $R$ and hence the
limit of $R \rightarrow 0$ is not taken.  This is particularly
important  for those systems where the scaling is significantly
different for small and large $R$ values [24].  The method
of computing the time delay $\tau$ in this scheme need not
be optimal for some specific data sets.  
Nevertheless,  under certain real life conditions,  
like when the number of
data sets are many and/or when a change in the state of
a system needs to be evaluated quickly and/or when qualitative
differences between time series needs to be estimated,  
the present scheme is recommended as a useful tool to compute the correlation 
dimension (and compare with surrogates) without non-algorithmic interventions.

KPH and GA acknowledge the hospitality and facilities in
IUCAA.  The authors thank the Department of Epileptology,  University of 
Bonn, 
for making the human brain EEG data available on their website.

\end{document}